\documentclass[aps,prl,superscriptaddress,twocolumn,showpacs]{revtex4-1}
\usepackage{amsmath,graphicx,bm}
\usepackage[colorlinks,linkcolor=blue,anchorcolor=green,citecolor=red]{hyperref}

\begin{document}
\title{Dynamic Feedback in Ferromagnet--Spin Hall Metal Heterostructures}

\author{Ran Cheng}
\affiliation{Department of Physics, Carnegie Mellon University, Pittsburgh, Pennsylvania 15213, USA}

\author{Jian-Gang Zhu}
\affiliation{Department of Electrical and Computer Engineering, Carnegie Mellon University, Pittsburgh, Pennsylvania 15213, USA}

\author{Di Xiao}
\affiliation{Department of Physics, Carnegie Mellon University, Pittsburgh, Pennsylvania 15213, USA}


\begin{abstract}

In ferromagnet/normal metal heterostructures, spin pumping and spin-transfer torques are two reciprocal processes that occur concomitantly. Their interplay introduces a dynamic feedback effect interconnecting energy dissipation channels of both magnetization and current. By solving the spin diffusion process in the presence of the spin Hall effect in the normal metal, we show that the dynamic feedback gives rise to: (i) a nonlinear magnetic damping that is crucial to sustain uniform steady-state oscillations of a spin Hall oscillator at large angles. (ii) a frequency dependent spin Hall magnetoimpedance that reduces to the spin Hall magnetoresistance in the dc limit.

\end{abstract}

\maketitle

\textit{Introduction.}---A central concept in modern spintronics is the emergence of artificial electromagnetics due to the interplay between magnetization dynamics and electron transport. For instance, when an electron spin adiabatically follows a slowly-varying magnetization, its wave function acquires a geometric phase changing with time. This phase resembles a time-varying magnetic flux and produces a spin motive force (SMF) according to the Faraday effect~\cite{ref:SMF,ref:Shengyuan}. As a feedback, electrons driven by SMFs react on the magnetization via the spin-transfer torque (STT)~\cite{ref:Slonczewski,ref:Beger,ref:Bazaliy,ref:STT}, which enhances the magnetic damping~\cite{ref:Zhang} to hinder the magnetization dynamics that causes the SMF. In a reciprocal sense, if a magnetic texture is driven into motion by a current, it in turn exerts SMFs on the electrons, modifying the electrical resistivity~\cite{ref:JDZang,ref:Schulz}. The feedback mechanism persists even in the presence of thermal and mechanical forces~\cite{ref:heatpump}, or when spin-orbit interactions are strong~\cite{ref:SOMF}. These examples constituent a general manifestation of Lenz's law in artificial electromagnetic, which states that a motive force induction always opposes the change of flux that causes the motive force, and vice versa~\cite{ref:Lenz}. In generic settings, Lenz's law imposes a universal rule on how a process can be affected by its converse: feedback should be negative, otherwise energy is not conserved.

\begin{figure}[t]
	\centering
	\includegraphics[width=\linewidth]{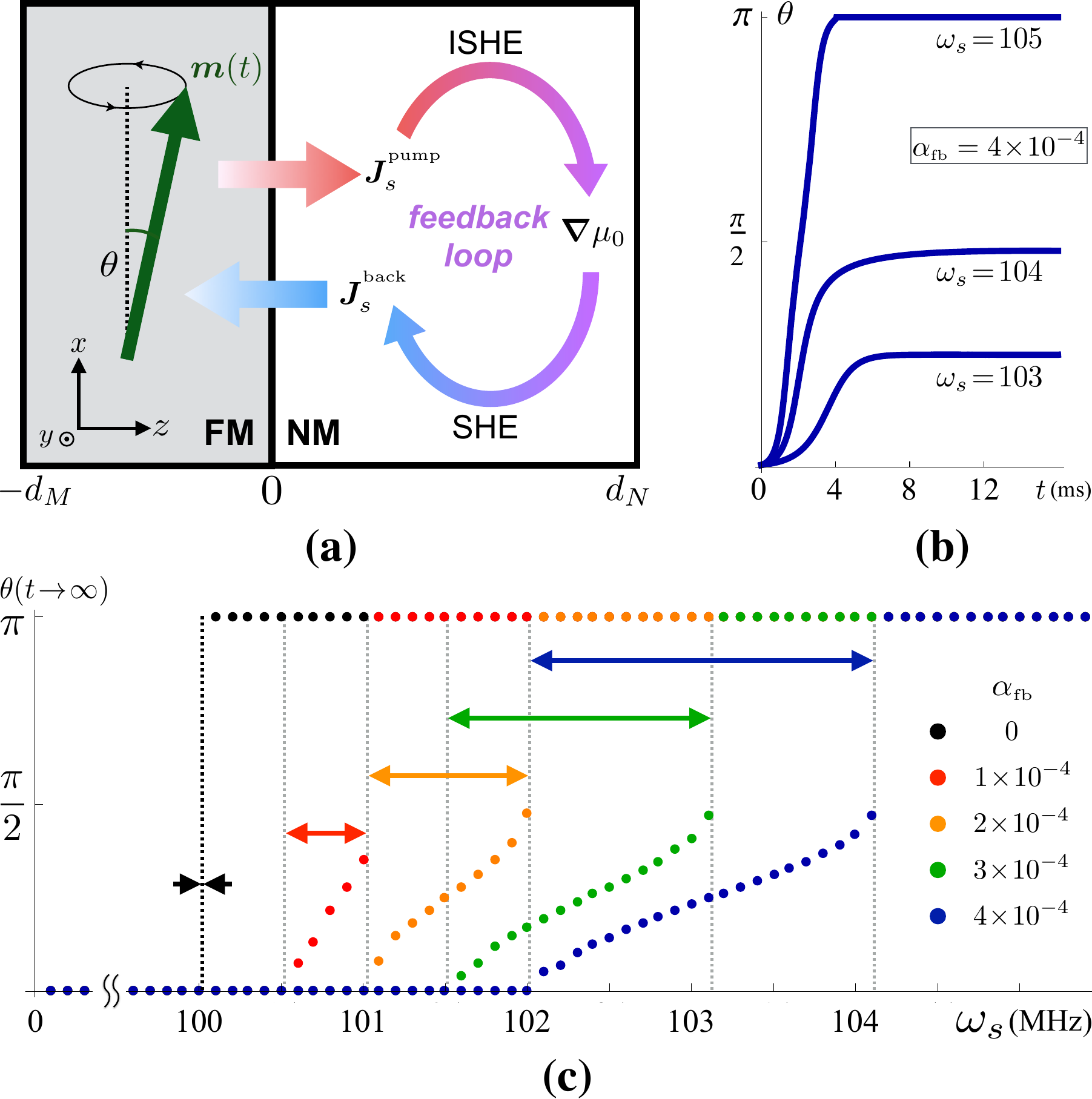}
	\caption{(Color online) (a) In a FM/NM bilayer, spin pumping and spin backflow are connected by the SHE and its inverse process. (b) and (c): Simulations of a spin Hall nano-oscillator in the presence of the feedback-induced nonlinear damping $\alpha_{\mbox{\tiny fb}}$, with $\gamma H=10$GHz, $d_m=1$nm, $\alpha_0+\alpha_{\mbox{\tiny sp}}=0.01$, and other parameters taken from Ref.~\cite{ref:YIGexp}. The STT strength $\omega_s$ is scaled in Megahertz.}
	\label{fig:loop}
\end{figure}

In all known phenomena so far, electrons and the magnetization couple \textit{locally} in the bulk~\cite{ref:Hydro}. Therefore, one is able to eliminate either the magnetization dynamics or the electron motion at arbitrary locations to derive the feedback renormalization of various response coefficients. In ferromagnet (FM)/normal metal (NM) heterostructures, however, \textit{nonlocal} effects arise because conduction electrons and magnetization reside in different materials and couple only at the interface. In this scenario, a precessing FM can pump spin current into the NM~\cite{ref:spinpumping,ref:Kajiwara}, which subsequently experiences a backflow and reacts on the FM via the STT~\cite{ref:spinbattery}. The combined effect of spin pumping and the backflow-induced STT renormalizes the interfacial transverse conductance~\cite{ref:backflow1}, and captures a static feedback effect involving nonlocal processes. However, recent experiments showed that the spin Hall effect (SHE) in the NM can drastically modify the dynamical behavior of the entire heterostructure~\cite{ref:Ralph,ref:Sinova}. Taking into account the SHE, spin pumping and spin backflow are also connected via the combined effect of the SHE and its inverse process, which forms a feedback loop as illustrated in Fig.~\ref{fig:loop}(a). This additional feedback mechanism, proportional to $\theta_s^2$ ($\theta_s$ is the spin Hall angle), was completely ignored in previous studies~\cite{ref:backflow3,ref:JX}. Nevertheless, the recently discovered spin Hall magnetoresistance (SMR)~\cite{ref:SMRexp1,ref:SMRexp2,ref:SMRtheory} reveals that physics at the $\theta_s^2$ level is essential to the electron transport. As the reported spin Hall angle $\theta_s$ is getting larger~\cite{ref:Niimi,ref:TI}, it is tempting to ask whether a feedback effect proportional to $\theta_s^2$ can alter the magnetization dynamics or the electron transport in a qualitative way.

In this Letter, we show that our proposed feedback mechanism manifests as a novel \textit{nonlinear} damping effect in the FM dynamics. It enables \textit{uniform} steady-state auto-oscillations of a spin Hall oscillator by preventing it from growing into magnetic switching. If our proposed feedback effect is ignored, however, auto-oscillations are possible only for spin-valves without the participation of the SHE~\cite{ref:Zangwill}, for materials with strong dipolar interactions~\cite{ref:Rezende}, or for spatially localized solitons in a FM/NM heterostructure~\cite{ref:Bullet,ref:Urazhdin}. In a reciprocal sense, we show that the feedback loop also gives rise to a spin Hall magnetoimpedance in the electron transport which reduces to the observed SMR in the dc limit.

\textit{Formalism.}---Consider a FM/NM bilayer structure as shown in Fig.~\ref{fig:loop}(a), where the layer thicknesses are $d_M$ and $d_N$, respectively. The coordinate system is chosen such that the magnetization direction at rest is along $x$, and the interface normal is along $z$. We assume that the FM is insulating (\textit{e.g.}, YIG), but the essential physics remains valid for a conducting FM. Let $\mu_0/2$ be the electrochemical potential and $\bm{\mu}$ the vector of spin accumulation in the NM, so by Ohm's law the charge current density is $J^c_i=-\frac{\sigma}{2e}[\partial_i\mu_0 + \theta_s\varepsilon_{ijk}\partial_j\mu_k]$, and the spin current density is $J^s_{ij}=-\frac{\sigma}{2e}[\partial_i\mu_j - \theta_s\varepsilon_{ijk}\partial_k\mu_0]$ with $i$ the transport direction and $j$ the direction of spin polarization. In our device geometry, only the spin current flowing along $z$-direction is relevant, thus we assume $\bm{\mu}=\bm{\mu}(z,t)$. Correspondingly, the spin current density reduces to a vector $\bm{J}_s$; we scale it in the same unit as the charge current density $\bm{J}_c$. The electron and spin dynamics in the NM are then described by three equations
\begin{align}
 \frac{\partial\bm{\mu}}{\partial t}&=D\frac{\partial^2\bm{\mu}}{\partial z^2}-\frac{1}{\tau_{\mathrm{sf}}}\bm{\mu}\ , \label{eq:spindiff} \\
 \bm{J}_c&=-\frac{\sigma}{2e}\left[ \bm{\nabla}\mu_0 + \theta_s\hat{\bm{z}}\times\frac{\partial\bm{\mu}}{\partial z} \right]\ , \label{eq:jc} \\
 \bm{J}_s&=-\frac{\sigma}{2e}\left[ \frac{\partial\bm{\mu}}{\partial z} + \theta_s\hat{\bm{z}}\times\bm{\nabla}\mu_0 \right]\ , \label{eq:js}
\end{align}
where $D$ is the diffusion constant, $\tau_{\mathrm{sf}}$ is the spin-flip relaxation time, $\sigma$ is the conductivity, $e$ is the electron charge, and $\theta_s$ is the spin Hall angle.

To solve the spin accumulation $\bm{\mu}$, we assume that the charge current density $\bm{J}_c$ is an applied dc charge current density which is fixed by external circuit. It only supplies a constant drive to the system but does not participate in the feedback process. To make it more specific, if we instead consider a constant voltage drive $\bm{\nabla}\mu_0=$const., then $\bm{J}_c$ and $\bm{\nabla}\mu_0$ will switch roles in Eq.~\eqref{eq:jc} and~\eqref{eq:js}. In other words, either $\bm{J}_c$ or $\mu_0$ must depend on $z$ while the other is uniform in space. In the following, we focus on a constant current drive condition and allow $\mu_0=\mu_0(z)$. In addition, we have two boundary conditions~\cite{ref:JX}: $\bm{J}_s(d_{N})=0$ and
\begin{align}
 \bm{J}_{s0}\equiv\bm{J}_s(0)=\frac{G_r}{e}\left[ \bm{m}\times(\bm{m}\times\bm{\mu}_{s0})+\hbar\bm{m}\times\dot{\bm{m}} \right]\ ,\label{eq:js0}
\end{align}
where we used the macrospin model and $\bm{m}$ is the unit vector of the magnetization. $\bm{\mu}_{s0}$ stands for $\bm{\mu}(0)$ and $G_r$ is the real part of the areal density of the spin-mixing conductance (the imaginary part $G_i$ is neglected since $G_i\ll G_r$~\cite{ref:note}). The $\bm{m}\times(\bm{m}\times\bm{\mu}_{s0})$ and $\hbar\bm{m}\times\dot{\bm{m}}$ terms represent STT and spin pumping, respectively. They are two fundamental ingredients bridging the electron (spin) transport in the NM with the FM. Due to spin conservation, the spin current density $\bm{J}_{s0}$ must be added to the Landau-Lifshitz-Gilbert (LLG) equation~\cite{ref:spinbattery,ref:backflow1}
\begin{align}
	\frac{d\bm{m}}{dt}=\gamma\bm{\bm{H}}_{\mbox{\tiny eff}}\times\bm{m}+\alpha_{0}\bm{m}\times\frac{\partial\bm{m}}{\partial t}+\frac{\hbar\gamma}{2eM_sd_M}\bm{J}_{s0}\ ,
	\label{eq:LLG}
\end{align}
where $\gamma$ is the gyromagnetic ratio, $\hbar$ is the reduced Planck constant, $M_s$ is the saturation magnetization, $\alpha_0$ is the Gilbert damping constant, and $\bm{\bm{H}}_{\mbox{\tiny eff}}$ is the effective magnetic field. 

The typical frequency $\omega$ of magnetization oscillation is much smaller than the spin relaxation rate $1/\tau_{\mathrm{sf}}$. As a result, the spin accumulation $\bm{\mu}(z,t)$ adapts to the instantaneous magnetization and is kept quasi-equilibrium~\cite{ref:backflow3}, and the spin dynamics described by Eq.~\eqref{eq:spindiff} reduces to a stationary spin diffusion process at every instant of time. Retaining to the $\theta_s^2$ order, Eq.~\eqref{eq:spindiff} is solved as
\begin{align}
 \bm{\mu}(z)=&\ \theta_{s}\frac{2e\lambda}{\sigma}\hat{\bm{z}}\times\bm{J}_c\frac{\sinh\frac{2z-d_{N}}{2\lambda}}{\cosh\frac{d_{N}}{2\lambda}} \notag\\
 & +\frac{2e\lambda}{\sigma} \left[\bm{J}_{s0}+\theta_{s}^{2}\hat{\bm{z}}\times\left(\hat{\bm{z}}\times\bm{J}_{s0}\right)\right]\frac{\cosh\frac{z-d_{N}}{\lambda}}{\sinh\frac{d_{N}}{\lambda}} \ , \label{eq:mu}
\end{align}
where $\lambda=\sqrt{D\tau_{\mathrm{sf}}}$ is the spin diffusion length. Here, we suppress the $t$ variable in $\bm{\mu}(z)$ since its time dependence simply originates from $\bm{J}_c$ and $\bm{J}_{s0}$. Combining Eq.~\eqref{eq:spindiff}---Eq.~\eqref{eq:mu}, we can either eliminate the electron degrees of freedom to derive an effective magnetization dynamics, or eliminate the time derivative of the magnetization ($\dot{\bm{m}}$) to get an effective magneto-transport of the electrons. These operations invoke our proposed dynamic feedback mechanism to the FM/NM heterostructure.

\textit{Nonlinear damping.}---Our goal is to express the spin current density flowing across the interface $\bm{J}_{s0}$ in terms of the magnetization $\bm{m}(t)$, by which the LLG Eq.~\eqref{eq:LLG} will no longer involve any electron degree of freedom except $\bm{J}_c$. To this end, we combine Eq.~\eqref{eq:js0} and Eq.~\eqref{eq:mu} for $z=0$, and obtain two convoluted relations of $\bm{J}_{s0}$ and $\bm{\mu}_{s0}$. By means of iterations truncating at $\theta_s^2$ order, we can solve $\bm{J}_{s0}$ as a function of $\bm{J}_c$, $\bm{m}(t)$ and its time derivative. Then we insert this $\bm{J}_{s0}$ into Eq.~\eqref{eq:LLG}, which yields the effective magnetization dynamics
\begin{align}
	\frac{d\bm{m}}{dt}=&\ \gamma\bm{\bm{H}}_{\mbox{\tiny eff}}\times\bm{m} +\omega_s\bm{m}\times\left[(\hat{\bm{z}}\times\hat{\bm{j}_c})\times\bm{m}\right]  \notag\\
	&+(\alpha_0+\alpha_{\mbox{\tiny sp}})\bm{m}\times\frac{\partial\bm{m}}{\partial t} \notag\\
	&+\alpha_{\mbox{\tiny fb}}\left( m_{z}^{2}\bm{m}\times\frac{\partial\bm{m}}{\partial t}+\frac{\partial m_z}{\partial t}\bm{m}\times\hat{\bm{z}} \right) \ , \label{eq:centralresult}
\end{align}
where $\hat{\bm{j}_c}$ is the unit vector of $\bm{J}_c$ and
\begin{align}
 \omega_s&=\theta_sJ_c\frac{\hbar\gamma}{eM_sd_M}\frac{\lambda G_r\tanh\frac{d_N}{2\lambda}}{\sigma+2\lambda G_r\coth\frac{d_N}{\lambda}} \label{eq:omegas}
\end{align}
is the strength of the STT (driven by $\bm{J}_c$) scaled in the frequency dimension. The two damping coefficients are
\begin{align}
	\alpha_{\mbox{\tiny sp}}&=\frac{\hbar^2\gamma}{2e^2M_sd_M}\frac{\sigma G_r}{\sigma+2\lambda G_r\coth\frac{d_N}{\lambda}}\ , \label{eq:alpha} \\
	\alpha_{\mbox{\tiny fb}}&=\theta_{s}^2\frac{\hbar^2\gamma}{e^2M_sd_M} \frac{\sigma\lambda G_r^2\coth\frac{d_N}{\lambda}}{(\sigma+2\lambda G_r\coth\frac{d_N}{\lambda})^2}\ . \label{eq:alphaFB}
\end{align}
Here, $\alpha_{\mbox{\tiny sp}}$ describes the conventional enhanced damping from spin pumping with the spin backflow effects taken into account~\cite{ref:spinbattery,ref:backflow1,ref:backflow3,ref:JX}; it is independent of the SHE. By contrast, the $\alpha_{\mbox{\tiny fb}}$ term is completely new. It reflects the dynamic feedback realized by virtue of the combined effect of the SHE and its inverse process as schematically shown in Fig.~\ref{fig:loop}(a). From Eq.~\eqref{eq:centralresult}, we see that this novel damping term is \textit{nonlinear} in $\bm{m}_{\perp}$---the component of $\bm{m}$ transverse to the effective field $\bm{\bm{H}}_{\mbox{\tiny eff}}$, whereas the Gilbert damping term is linear in $\bm{m}_{\perp}$.

The feedback-induced nonlinear damping effect can be understood in an intuitive way. If the magnetization precession is getting larger, it will trigger a chain reaction: first the pumped spin current $\bm{J}_{s0}$ increases, then the spin diffusion becomes stronger (\textit{i.e.}, $|\partial_z\bm{\mu}|$ gets larger). This will necessarily lead to a larger emf $\bm{\nabla}\mu_0$ in the NM according to Eq.~\eqref{eq:jc}, as we have fixed the current density $\bm{J}_c$. The change of the emf will eventually feed back into $\bm{J}_{s0}$ according to Eq.~\eqref{eq:js}, limiting its further growth. As a consequence, the growing magnetization precession is inhibited. If we draw an analogy between the magnetization oscillation and an electric motor, the feedback loop realizes an effective back emf induction preventing the electric motor from rotating faster.

\textit{Example.}---We demonstrate the physical significance of the nonlinear damping effect in a current-driven spin Hall nano-oscillator. Consider that the magnetization is polarized by a magnetic field $\bm{H}=H\hat{\bm{x}}$, and is driven by a dc current density $\bm{J}_c=J_c\hat{\bm{y}}$. To determine the threshold of auto-oscillation excitation, we assume that $\bm{m}(t)=\hat{\bm{x}}+\bm{m}_{\perp}e^{i\omega t}$ where $\bm{m}_{\perp}=m_y+im_z$ and $|\bm{m}_{\perp}|\ll1$, and regard $\omega$ as a complex frequency where the imaginary part represents the damping. Inserting the above Ansatz into Eq.~\eqref{eq:centralresult} and setting $\mathrm{Im}[\omega]=0$ yield the threshold STT strength: $\omega_s^{\mathrm{th}}=(\alpha_0+\alpha_{\mbox{\tiny sp}}+\alpha_{\mbox{\tiny fb}}/2)\gamma H$, which can be converted to a threshold current density $J_c^{\mathrm{th}}$ by Eq.~\eqref{eq:omegas}. In the beyond threshold regime, $J_c>J_c^{\mathrm{th}}$, $\bm{m}_{\perp}$ starts to grow exponentially in time. If $\alpha_{\mbox{\tiny fb}}=0$, however, the growth will ultimately evolve into a magnetic switching. This is because the driving STT and the Gilbert damping are both linear in $\bm{m}_{\perp}$ so that if the former overcomes the latter it wins at arbitrary angles $\theta=\arcsin m_{\perp}$. As a result, whenever a spontaneous motion is triggered, its amplitude will grow indefinitely. The only way to enable stable oscillation at an intermediate configuration is to make the overall damping grow faster than the driving STT with an increasing $\bm{m}_{\perp}$, \textit{i.e.}, the damping has to be nonlinear in $\bm{m}_{\perp}$. By doing so, the amplitude growth will terminate at an angle where the two competing mechanisms compensate each other, and a steady-state oscillation is realized there. The feedback-induced nonlinear damping effect just fulfills this need. From the perspective of dynamical stability, after a steady-state oscillation is achieved, the damping (the STT) will dominate again if the angle $\theta$ is getting larger (smaller) so that the magnetization will be dragged back. We mention in passing that our proposed feedback mechanism is not exclusive to FMs, but applies to antiferromagnets as well when integrated with the SHE~\cite{ref:AFoscillator}.

To justify the above prediction, we perform a series of numerical simulations. From Eq.~\eqref{eq:alphaFB}, we know that a smaller (larger) $d_M$ ($d_N$) leads to a larger $\alpha_{\mbox{\tiny fb}}$. Consider an YIG/Pt bilayer structure with $d_M$ a few nanometers and $d_N\gg\lambda$, and other material parameters taken from a recent experiment~\cite{ref:YIGexp}, then $\alpha_{\mbox{\tiny fb}}$ is estimated to be of order $10^{-4}$, comparable to the intrinsic Gilbert damping $\alpha_0$ in YIG. Assuming $\gamma H=10$GHz, $\alpha_0+\alpha_{\mbox{\tiny sp}}=0.01$ and $\alpha_{\mbox{\tiny fb}}=4\times10^{-4}$, we plot in Fig.~\ref{fig:loop}(b) the precession angle $\theta$ as a function of time for three different STT strengths $\omega_s$ (scaled in Megahertz). We also plot in Fig.~\ref{fig:loop}(c) the terminal angle $\theta(t\!\rightarrow\!\infty)$ as a function of $\omega_s$ for four different values of $\alpha_{\mbox{\tiny fb}}$. In Fig.~\ref{fig:loop}(c), two features are evident: (i) a larger $\omega_s$ (larger driving current density $J_c$) results in a larger terminal angle, but at sufficiently large $\omega_s$, the oscillator inevitably undergoes a magnetic switching. (ii) a lager $\alpha_{\mbox{\tiny fb}}$ (stronger feedback) widens the window of steady-state oscillations. These results have justified that the nonlinear damping effect described by Eq.~\eqref{eq:centralresult} can indeed sustain stable oscillations. 

Next we comment on several side-effects that could potentially obscure the observation of our predictions. First, if the FM film is too thin, the dipolar interaction might not be negligible, which can cause magnon-magnon scattering that provides a different nonlinearity to bound a spontaneous excitation from blowing up~\cite{ref:Rezende}. When the dipolar effect dominates, the nonlinear damping effect is undermined. However, if the magnon-magnon scattering is negligible and the dipolar effect can be approximated by a hard-axis anisotropy, the nonlinear damping effect should still be observable, but the steady-state precession will become elliptical. Second, in existing realizations of spin Hall oscillators such as Ref.~\cite{ref:Urazhdin}, a point-contact is often used. A known fact about such experimental setup is that it can easily excite the spatially localized mode (soliton)~\cite{ref:Bullet} rather than a uniform oscillation. Finally, a steady-state oscillation seems to be possible if we apply the driving current density $\bm{J}_c$ parallel to $\bm{m}$ (so the spin accumulation is perpendicular to $\bm{m}$ due to the device geometry). However, in that case the oscillation cannot be regarded as an auto-oscillation of the \textit{eigenmode} with a fixed frequency. Instead, the magnetization undergoes consecutive precessional switching with a frequency proportional to $J_c$~\cite{ref:ultrafast}. While this still forms an oscillator, it is not able to directly verify the physical significance of our nonlinear damping effect.

\textit{Spin Hall magnetoimpedance.}---As a reciprocal effect, the dynamic feedback also affects the electron transport. If we apply an ac current density $\bm{J}_c(t)=\tilde{\bm{J}}_ce^{i\omega t}$ to an FM/NM heterostructure longitudinally, the SHE will drive the magnetization precession via the STT, which in turn can pump spin current back into the NM and renormalize the resistivity by means of the inverse SHE. This is analogous to an ac electric motor accommodating the counteractive motive force induced by the simultaneous dynamotor effect. Although the feedback received by an ac current drive has been studied from the angle of STT-induced ferromagnetic resonance~\cite{ref:Ralph,ref:SMI,ref:STFMR,ref:plus}, we explore its phenomenology from the feedback perspective, which is conceptually advanced and reveals new insights.

\begin{figure}[t]
	\centering
	\includegraphics[width=0.95\linewidth]{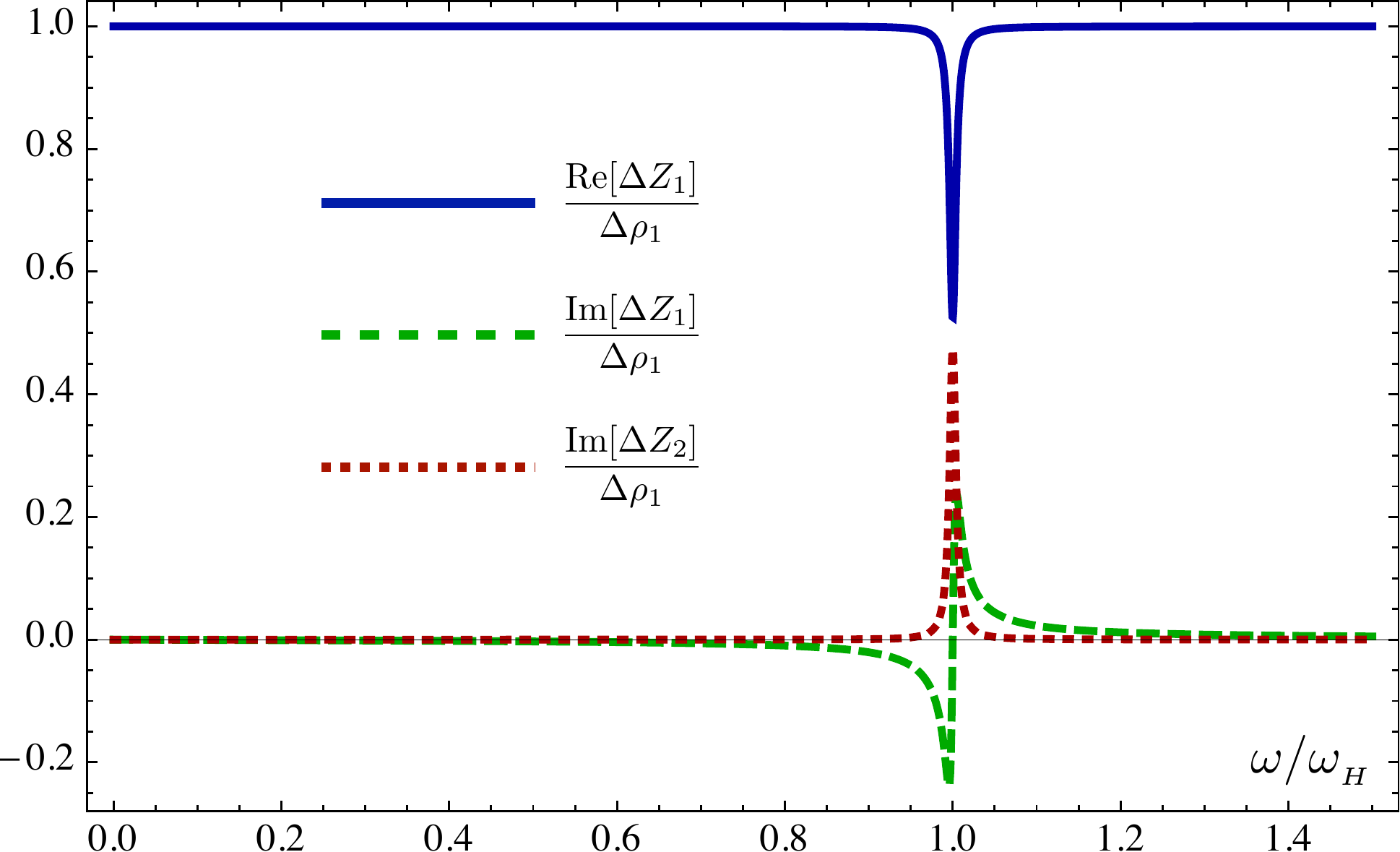}
	\caption{(Color online) Frequency dependence of the spin Hall magnetoimpedance scaled by $\Delta\rho_1\equiv\lim_{\omega\rightarrow0}\mathrm{Re}[\Delta Z_1]$. $\mathrm{Re}[\Delta Z_2]$ is not shown since $\mathrm{Re}[\Delta Z_2]=\mathrm{Im}[\Delta Z_1]$. The plot is based on an YIG/Pt structure~\cite{ref:YIGexp} with $\alpha_0=2.3\times10^{-4}$, $d_M=1$nm, and $d_N\gg\lambda$.}
	\label{fig:smi}
\end{figure}

Consider that the magnetization $\bm{m}(t)$ is oscillating uniformly around an applied magnetic field $\bm{H}=H\hat{\bm{h}}$. By performing a Fourier transformation, we can rewrite Eq.~\eqref{eq:LLG} in the frequency domain (where quantities are capped with tildes) and obtain
\begin{align}
 \tilde{\bm{m}}_\perp=\frac{\hbar\gamma}{2eM_sd_F}\frac{ i\omega\tilde{\bm{J}}_{s0}+\left(\omega_H+i\alpha_0\omega\right)\hat{\bm{h}}\times\tilde{\bm{J}}_{s0} }{(\omega_H+i\alpha_0\omega)^2-\omega^2}\ , \label{eq:mperp}
\end{align}
where $\omega_H=\gamma H$. Combining Eq.~\eqref{eq:mperp} with the spin current density flowing through the interface [Eq.~\eqref{eq:js0}], the spin accumulation [Eq.~\eqref{eq:mu}], and Ohm's law [Eq.~\eqref{eq:jc}], we are able to solve the (spatially) averaged electric field $\tilde{\bm{E}}\equiv-\frac{1}{2ed_{N}}\int_{0}^{d_N}\bm{\nabla}\tilde{\mu}_0dz$. Choosing the in-plane coordinates such that $\tilde{\bm{J}}_c=\tilde{J}_c\hat{\bm{x}}$, we obtain
\begin{subequations}
	\begin{align}
	\tilde{E}_x&=[\rho+\Delta\rho_0+\Delta Z_1(\omega)(1-h_y^2)]\tilde{J}_c\ ,\\
	\tilde{E}_y&=[\Delta Z_1(\omega)h_xh_y+\Delta Z_2(\omega)h_z]\tilde{J}_c\ ,
	\end{align}
\end{subequations}
where $\rho=1/\sigma$ is the intrinsic bulk resistivity of the NM without including any feedback effect. Here, the spin Hall magnetoimpedance (SMI) consists of three distinct contributions: one frequency independent (dc) component $\Delta\rho_0/\rho=-\theta_{s}^{2}\frac{2\lambda}{d_{N}}\tanh\frac{d_{N}}{2\lambda}$, and two frequency dependent components
\begin{align}
\frac{\Delta Z_1(\omega)}{\rho}=&\ \theta_{s}^{2}\frac{\lambda^2\rho G_r}{d_N}\frac{(1+U+\mathcal{P}_\omega)\tanh^{2}\frac{d_{N}}{2\lambda}}{(1+U+\mathcal{P}_\omega)^2+\mathcal{Q}^2_\omega}\ , \label{eq:Z1}\\
\frac{\Delta Z_2(\omega)}{\rho}=&-\theta_{s}^{2}\frac{\lambda^2\rho G_r}{d_N}\frac{\mathcal{Q}_\omega\tanh^{2}\frac{d_{N}}{2\lambda}}{(1+U+\mathcal{P}_\omega)^2+\mathcal{Q}^2_\omega}\ , \label{eq:Z2}
\end{align}
where $U=2\rho G_r\lambda\coth\frac{d_{N}}{\lambda}$ and
\begin{subequations}
\label{eq:PQ}
 \begin{align}
  \mathcal{P}_\omega=\frac{\hbar^2\gamma G_r}{2e^2 M_s d_F}\frac{i\omega(\omega_{H}+i\alpha_{0}\omega)}{\left(\omega_{H}+i\alpha_{0}\omega\right)^{2}-\omega^{2}}\ , \\
  \mathcal{Q}_\omega=\frac{\hbar^2\gamma G_r}{2e^2 M_s d_F}\frac{\omega^2}{\left(\omega_{H}+i\alpha_{0}\omega\right)^{2}-\omega^{2}}\ .
 \end{align}
\end{subequations}
In the dc limit $\omega\rightarrow0$, the above results reduce to the recently discovered SMR~\cite{ref:SMRexp1,ref:SMRexp2,ref:SMRtheory}. Eq.~\eqref{eq:Z1} and~\eqref{eq:Z2} give us a relation $\mathrm{Re}[\Delta Z_2]=\mathrm{Im}[\Delta Z_1]$, which will break down if the imaginary part of the spin-mixing conductance $G_i$ is included in our calculation~\cite{ref:note}. In Fig.~\ref{fig:smi}, we plot $\Delta Z_1(\omega)$ and $\Delta Z_2(\omega)$ as functions of the frequency $\omega$ with all quantities scaled by the longitudinal SMR $\Delta\rho_1$. Fig.~\ref{fig:smi} shows that a pronounced deviation of the SMI from the SMR takes place only in the vicinity of the STT-induced ferromagnetic resonance. This deviation, according to Eqs.~\eqref{eq:Z1}, \eqref{eq:Z2}, and~\eqref{eq:PQ}, scales roughly as $d_N/d_M$ when $d_N$ is small. In a recent measurement~\cite{ref:SMI}, the observed deviation of the SMI from the SMR is negligibly small, probably because their FM is too thick ($d_M$=55nm) while the NM is too thin ($d_N$=4nm).

The Oersted field generated by $\bm{J}_c$ is also responsible for the SMI~\cite{ref:STFMR}. But one can distinguish the feedback contribution and the Oersted field contribution from the symmetry pattern of SMI with respect to $(\omega-\omega_H)$. For instance, $\mathrm{Re}[\Delta Z_1]$ due to the dynamic feedback is symmetric around $\omega_H$, whereas it becomes antisymmetric when the Oersted field is dominating. The relative ratio of the two contributions depends on the NM thickness $d_N$. For fixed dc current $J=J_cd_N$, the Oersted field is fixed, but the STT is basically proportional to $d_N$ for $d_N>\lambda$ as shown by Eq.~\eqref{eq:omegas}. Therefore, to observe an overwhelming feedback contribution, both the NM and the FM should be thin (while keeping $d_N>\lambda$). This feature has been verified in a recent experiment~\cite{ref:plus}.

\begin{acknowledgments}
 R.C. and D.X. are indebted to A. Brataas for insightful discussions and a detailed check of the derivations. The authors are grateful to J. Xiao and M. W. Daniels for important comments. This study was supported by the U.S. Department of Energy, Office of BES, Division of MSE under Grant No.~DE-SC0012509.
\end{acknowledgments}


\begin{thebibliography}{20}
\bibitem{ref:note} If one keeps $G_i$, the central result Eq.~\eqref{eq:alphaFB} does not change much: the factor $G_r^2/(\sigma+2\lambda G_r\coth d_N/\lambda)^2$ will become $|\mathcal{G}|^2/|\sigma+2\lambda \mathcal{G}\coth d_N/\lambda|^2$ with $\mathcal{G}=G_r+iG_i$.
\end{thebibliography}
\end{document}